\documentclass[9pt]{extarticle}
\usepackage     [utf8]                  {inputenc}
\usepackage     [T1]                    {fontenc}
\usepackage                             {color}
\usepackage                             {amsmath}
\usepackage                             {graphicx}
\usepackage     [german,english]               {babel}
\usepackage                             {hyperref}
\usepackage{siunitx}
\usepackage{graphicx}
\usepackage{appendix}
\usepackage{lscape}
\usepackage{physics}
\usepackage[colaction]{multicol}
\usepackage{setspace} 
\usepackage{authblk}
\usepackage{xcolor,soul,framed,caption}
\colorlet{shadecolor}{yellow}
\usepackage[left=2cm,right=2cm,top=1cm,bottom=1.5cm,includeheadfoot]{geometry}
\title{Direct measurement of the $^3$He$^+$ magnetic moments}
\date{}
\author[1,*]{A. Schneider}
\author[1]{B. Sikora}
\author[1]{S. Dickopf}
\author[1]{M. M\"uller}
\author[1]{N. S. Oreshkina}
\author[1]{A. Rischka}
\author[1]{I. A. Valuev}
\author[2]{\newline S. Ulmer}
\author[3,4]{J. Walz}
\author[1]{Z. Harman}
\author[1]{C. H. Keitel}
\author[1]{A. Mooser}
\author[1]{K. Blaum}
\affil[1]{Max Planck Institute for Nuclear Physics, Saupfercheckweg 1, D-69117, Heidelberg, Germany}
\affil[2]{RIKEN, Ulmer Fundamental Symmetries Laboratory, 2-1 Hirosawa, Wako, Saitama, 351-0198, Japan}
\affil[3]{Institute for Physics, Johannes Gutenberg-University Mainz, Staudinger Weg 7, D-55099 Mainz, Germany}
\affil[4]{Helmholtz Institute Mainz, Staudingerweg 18, D-55128 Mainz, Germany}
\affil[*]{e-mail: antonia.schneider@mpi-hd.mpg.de}
\usepackage{amsmath,amssymb}
\usepackage{amsthm}
\usepackage{paralist}
\usepackage{algorithmic}
\usepackage{graphicx}
\usepackage{url}
\usepackage[babel,german=quotes]{csquotes}
\usepackage{simplewick}
\usepackage{slashed}
\usepackage{color}
\usepackage{braket}

\theoremstyle{definition}  
\theoremstyle{definition}  
\theoremstyle{remark}      
\theoremstyle{remark}      
\theoremstyle{remark}

\usepackage{lineno}

\begin{document}
\maketitle
\textbf{Helium-3 has nowadays become one of the most important candidates for studies in fundamental physics~\cite{intro2_3he, deuteronmass, superfluid}, nuclear and atomic structure~\cite{intro1_3he, 3he4heradius}, magnetometry and metrology~\cite{chupp} as well as chemistry and medicine~\cite{chemistry3He, medicine}. In particular, $^3$He nuclear magnetic resonance (NMR) probes have been proposed as a new standard for absolute magnetometry \cite{chupp, niki14}. This requires a high-accuracy value for the  $^3$He nuclear magnetic moment, which, however, has so far been determined only indirectly and with a relative precision of $12$ parts per billon (p.p.b.)~\cite{flowers93,neronov14}.
Here we investigate the $^3$He$^+$ ground-state hyperfine structure in a Penning trap to directly measure the nuclear \textit{g}-factor of $^3$He$^+$ $g'_I=-4.255\, 099\, 606\, 9(30)_{stat}(17)_{sys}$, the zero-field hyperfine splitting $E_\textnormal{HFS}^{\rm exp}=-8\, 665\, 649\, 865.77(26)_{stat}(1)_{sys}$~$\si{\hertz}$ and the bound electron \textit{g}-factor $g_e^\text{exp}=-2.002\, 177\, 415\, 79(34)_{stat}(30)_{sys}$. The latter is consistent with our 
theoretical value $g_e^\text{theo}=-2.002\, 177\, 416\, 252\, 23(39)$ based on parameters and fundamental constants from \cite{CODATA2018}.
Our measured value for the $^3$He$^+$ nuclear $g$-factor allows for the determination of the \textit{g}-factor of the bare nucleus $g_I=-4.255\, 250\, 699\, 7(30)_{stat}(17)_{sys}(1)_{theo}$ via our accurate calculation of the diamagnetic shielding constant~\cite{Yerokhin2012} $\sigma_{^3\textnormal{He}^+}=0.000\,035\,507\,38(3)$. This constitutes the first direct calibration for $^3$He NMR probes and an improvement of the precision by one order of magnitude compared to previous indirect results.
The measured zero-field hyperfine splitting improves the precision by two orders of magnitude compared to the previous most precise value~\cite{EHFS_1969} and enables us to determine the Zemach radius~\cite{Zemach1956} to $r_Z=2.608(24)~\si{\femto\meter}$.
}

Precise and accurate measurements of fundamental properties of simple physical systems allow testing our understanding of nature and search for or constrain physics beyond the Standard Model of particle physics (SM). For example, the measurement of the hyperfine splitting of the $2s$ state of $^{3}\mathrm{He}^{+}$ \cite{PhysRevA.16.6} provides one of the most sensitive tests of the bound state QED theory \cite{Karshenboim2002} at low $Z$. However, measurements at improved precision inevitably demand an accurate description and better understanding of systematic effects to exclude experimental errors and a misinterpretation of the results. Prominent examples are inconsistencies in the masses of light ions, which are subject to re-examination in the context of the light ion mass puzzle~\cite{deuteronmass}. Moreover, a discrepancy between measurements of the hyperfine structure of $^{209}$Bi$^{82+,80+}$ and the predictions of the SM could be resolved by repeating NMR measurements to determine the nuclear magnetic moment of $^{209}$Bi~\cite{intro3_Bi,intro4_Bi}. Here we study the fundamental properties of another isotope with relevance for NMR, $^3$He. We report on the first direct determination of its nuclear magnetic moment, which is of utmost importance for absolute magnetometry as it constitutes the first direct and independent calibration of $^3$He NMR probes.

NMR probes, unlike superconducting quantum interference devices (SQUID) or giant magnetoresistance sensors (GMR), allow for measurements of the absolute magnetic field with high precision and $^3$He probes, in particular, offer a higher accuracy than standard water NMR probes~\cite{chupp}. Owing to the properties of noble gases, they require significantly smaller corrections due to systematic effects such as dependence on impurities, probe shape, temperature and pressure~\cite{niki14}. Moreover, the diamagnetic shielding $\sigma$ of the bare nuclear magnetic moment by the surrounding electrons is known more precisely for $^3$He than water samples, for which these contributions are only accessible by measurement. In case of atomic $^3$He, the factor $1-\sigma_{^3\textnormal{He}}$ correcting for the shielding by the two electrons has been calculated theoretically with a relative precision of $10^{-10}$~\cite{sighe}, where the uncertainty is given by neglected QED corrections. 
Thus, $^3$He probes have a wide variety of highly topical applications in metrology and field calibration in precision experiments such as the muon $g-2$ experiments at Fermilab and J-Parc~\cite{muon,japan}.
Up to now, however, the only measurements of the $^3$He nuclear magnetic moment are based on comparisons of the NMR frequency of $^3$He to that of water or molecular hydrogen~\cite{neronov78, flowers93, neronov14} and are limited to $12$ p.p.b. due to the uncertainty of the shielding factor of the protons in water.

We have constructed a new experiment that allows for the direct measurement of the $^3$He nuclear magnetic moment by investigating the hyperfine structure of a single $^3$He$^+$ ion in a Penning trap, providing the first direct and independent calibration of $^3$He NMR probes as well as improving the precision by a factor of $10$. The result establishes $^3$He probes as an independent standard for absolute and accurate magnetometry. Thus, it allows to calibrate water probes by measuring the ratio of water and $^3$He NMR frequencies, which enables the extraction of the shielded magnetic moment in water with a relative precision of 1 p.p.b. instead of 12 p.p.b..

\begin{figure}
    \includegraphics[width=89mm]{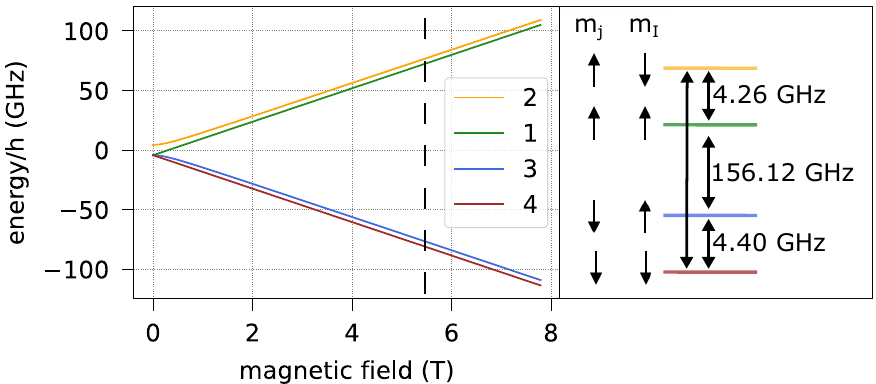}
    \caption{Breit-Rabi diagram of $^3$He$^+$.
The energies of the hyperfine states $E_1$, $E_2$, $E_3$ and $E_4$ are plotted as function of the magnetic field according to Eq.~\eqref{breitrabi_eq}. The arrows below $m_j$ and $m_I$ indicate the orientation with respect  to the magnetic  field of the total angular momentum of the electron $j=1/2$ and the nuclear spin $I=1/2$, which are antiparallel to the magnetic moments $\mu_e$ and $\mu_I$, respectively. The four double-headed arrows indicate the hyperfine transitions measured in this work. The transition frequencies given on the right side refer to the magnetic field in the Penning trap $B=5.7~\si{\tesla}$, which is marked in the plot by the black dashed line.} 
    \label{breitrabi_fig}
\end{figure}

In $^3$He$^+$, a splitting of the level structure arises due to the magnetic moment of the nucleus with nuclear spin $I=\frac{1}{2}$ interacting with the magnetic field generated by the orbiting electron. Investigating the level structure in an external magnetic field allows to extract the nuclear magnetic moment, which has been previously done with muonium \cite{PhysRevLett.82.711} and hydrogen \cite{PhysRevA.5.83}.
The combined hyperfine and Zeeman effect leads to a splitting of the $1s$ electronic ground state into four magnetic sublevels (Fig.~\ref{breitrabi_fig}), as described by the Breit-Rabi
formula \cite{Rabieq} up to first order perturbation theory in the magnetic field strength $B$:
\begin{equation}
E_{1,4} = \frac{E_{\rm HFS}}{4} \mp (\mu_IB + \mu_eB)\,, \quad
E_{2,3} = -\frac{E_{\rm HFS}}{4} \pm \frac{1}{2}\sqrt{E_{\rm HFS}^2 + 4(\mu_eB - \mu_IB)^2} \,. 
\label{breitrabi_eq}
\end{equation}
In these formulas, $E_{\rm HFS}<0$ is the hyperfine splitting at $B=0~\si{\tesla}$ and $\mu_e$ and $\mu_I$ are the spin magnetic moments of the electron and nucleus, respectively. However, at our experimental precision, second-order corrections of the above formula in $B$ have to be taken into account. These include the quadratic Zeeman shift, which is identical for all four levels involved and has therefore no influence on the transition frequencies, and the shielding correction~\cite{Moskovkin2006}. The latter effectively modifies the bare
nuclear $g$-factor $g_I$ to a shielded nuclear $g$-factor $g'_I=g_I(1-\sigma_{^3\textnormal{He}^+})$ of the ion, so that the magnetic moments in the equations above are related to the nuclear and electron $g$-factors via $\mu_{I}=g'_{I}\mu_N/2$ and $\mu_{e}=g_{e}\mu_B/2$. Here, $\mu_B=e\hbar/(2m_e)$ is the Bohr magneton, $\mu_N=e\hbar/(2m_p)$ is the nuclear magneton, $e$ is the elementary charge, $\hbar$ is the reduced Planck constant and $m_e$ and $m_p$ are the mass of the electron~\cite{electronm} and the proton~\cite{protonmass}.
In the current work, we combine measurements of four transition frequencies $(E_i(B)-E_j(B))/h$ to determine the three parameters $g'_I$, $g_e$ and $E_{\rm HFS}$, and additionally determine $g_e$, $E_{\rm HFS}$ and $\sigma_{^3\textnormal{He}^+}$ theoretically.
The latter is needed to calculate the bare nuclear $g$-factor from the measured $g'_I$. The theoretical and experimental results for $E_{\rm HFS}$, when combined with $g_I$, enable the extraction of a further nuclear parameter, namely, the Zemach radius characterizing the nuclear charge and magnetization distribution.

The interaction of the electron with the nuclear potential is taken into account by extending the free electron $g$-factor, in leading order corrected by the well-known Schwinger term~$\alpha/\pi$, with additional terms~\cite{Czarnecki00,Pachucki05}. The leading relativistic binding term then reads~\cite{Breit1928}
\begin{equation}
-g_{\rm Dirac}-2=\frac{4}{3}\left(\sqrt{1-(2\alpha)^2}-1\right)\,,
\end{equation}
which needs to be complemented with 1- to 5-loop QED binding corrections, as well as terms originating from the nucleus, i.e. the nuclear recoil term and nuclear structure effects. The numerical values of the contributing terms are given in the supplementary material. Our final result for the $g$-factor of the electron bound in $^3$He$^+$ is $g_{e}^{\rm theo}=-2.002\, 177\, 416\, 252\, 23(39)$, where the fractional accuracy is 0.15 parts per trillion (p.p.t.) and is dominantly limited by the uncertainty of $\alpha$ via the Schwinger term.

The theoretical contributions to the zero-field hyperfine splitting can be represented as~\cite{Beier00,Shabaev1994}
\begin{linenomath}
\begin{align}
E_{\mathrm{HFS}}=  \frac{4}{3} \alpha g_I \frac{m_e}{m_p} m_e c^2 (Z \alpha)^3 \mathcal{M} \left[ A(Z\alpha) + \delta_{\mathrm{FS}} + \delta_{\rm NP} +
\delta_{\mathrm{QED}} + \delta_{\rm \mu VP} + \delta_{\rm had VP} + \delta_{\rm ew} + \delta_{\mathrm{recoil}} \right]\,,
\end{align}
\end{linenomath}
where the relativistic factor is $A(Z\alpha) = {(2 \gamma + 1)}/{(\gamma (4 \gamma^2 -1))}$ with $\gamma= \sqrt{1-(Z\alpha)^2}$, and the mass prefactor is $\mathcal{M} = \left( 1 + \frac{m_e}{M_N} \right)^{-3}$ with the nuclear mass $M_N$. The $\delta$ correction terms in the above equation
denote finite nuclear size, nuclear polarization, QED, muonic and hadronic vacuum polarization, electroweak and nuclear recoil contributions, respectively. We evaluate these contributions as
described in the supplementary material and arrive at the theoretical hyperfine splitting of $E_{\mathrm{HFS}}^{\rm theo}=-8\,665\,701(19)$~kHz.
The calculation of the shielding constant is analogous to the theory of $g_e$ and $E_{\mathrm{HFS}}$ and further described in the supplementary material. The total value of this constant is
$\sigma_{^3\textnormal{He}^+}=0.000\,035\,507\,38(3)$, where the uncertainty is dominated by neglected higher order QED terms. This high accuracy, due to the low value of $Z\alpha$ and due to suppressed nuclear effects, enables an accurate extraction 
of the unshielded nuclear $g$-factor from the measured shielded $g$-factor.
\begin{figure}
    \centering
    \includegraphics[width=183mm]{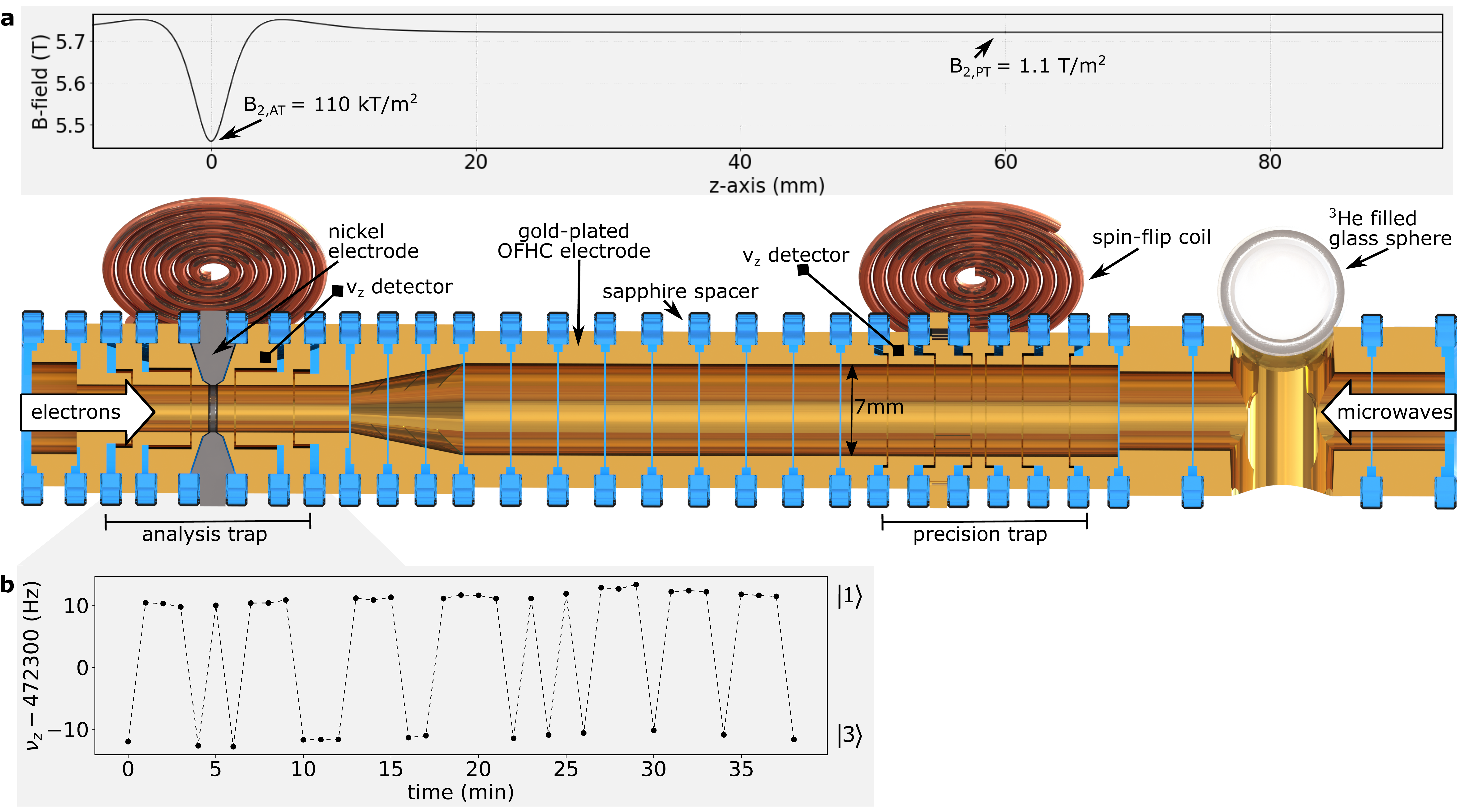}
    \caption{Schematic of the Penning-trap setup. A) Sectional view of the trap tower consisting of cylindrical electrodes and spatial variation of the magnetic field inside the trap tower along the $z$-axis. The insulation rings between the electrodes are depicted in blue, the copper electrodes yellow and the nickel electrode grey. All electrodes are gold-plated. The microwaves for driving spin-flips are introduced into the trap using the copper coils on the side of the trap and through a wave-guide from the top of the trap (white arrow) in case of the $4~\si{\giga\hertz}$ and $150~\si{\giga\hertz}$ transitions, respectively. 
    The second white arrow on the left side represents electrons from a field emission point used to ionize the atoms emitted by the $^3$He filled glass sphere. 
    The magnetic inhomogeneity in the analysis trap is spatially separated from the very homogeneous field in the precision trap by transport electrodes. 
    B) Axial frequency $\nu_z$ measured in the AT after resonantly driving the electronic transition $\ket{1}\leftrightarrow\ket{3}$. The dashed line serves to guide the eye. The frequency is higher by $22~\si{\hertz}$ when the ion is in state $\ket{1}$ compared to state $\ket{3}$. The same axial frequency shift can be observed when transitioning between states $\ket{2}$ and $\ket{4}$.}
    \label{fig_traps}
\end{figure}

In our single-ion Penning-trap experiment, we measure the transition frequencies between the hyperfine states in Eq.~\eqref{breitrabi_eq} and simultaneously the magnetic field, via the accurate determination of the free cyclotron frequency
\begin{equation}
    \nu_c=\frac{1}{2\pi}\frac{e}{m_{^3\textnormal{He}^+}}B,
    \label{nuc}
\end{equation}
where  $e/m_{^3\textnormal{He}^+}$ is the charge-to-mass ratio of the ion~\cite{CODATA2018}. 

The Penning trap setup shown in Fig.~\ref{fig_traps}A is placed in a $5.7~\si{\tesla}$ superconducting magnet and is in thermal contact with a liquid helium bath. 
In the analysis trap (AT) a nickel electrode creates a magnetic inhomogeneity that allows for the detection of the hyperfine state, as described below, but also limits the precision with which the ion's eigenfrequencies and the transition frequencies can be measured due to line-broadening \cite{doubletrapmethod}. These frequencies can be detected with high precision in a second trap, the precision trap (PT), which is separated by several transport electrodes from the AT so that the magnetic inhomogeneity is smaller by a factor of $10^{-5}$, see Fig.~\ref{fig_traps}A. 
A measurement cycle starts with determining the initial hyperfine state in the AT. The ion is then transported adiabatically to the PT, where the cyclotron frequency is first measured to determine the expected hyperfine transition frequency. It is afterwards measured again while a microwave excitation drives one of the four hyperfine transitions at a random frequency offset with respect to the expected resonance frequency. Whether a change of the hyperfine state occurred in the PT is then analyzed after transporting the ion back to the AT. This process is repeated several hundred times for each of the four transitions so as to measure the transition probability in the magnetic field of the PT as a function of the microwave frequency offset. 

The trap tower (Fig.~\ref{fig_traps}A) is enclosed by a trap chamber which is sealed off from the surrounding prevacuum to allow for ion storage times of several months \cite{lifetime_pbar}. Therefore, $^3$He can not be introduced to the trap by an external source but instead is released from the depicted SO$_2$ glass sphere, which is filled with $^3$He gas. Due to the strongly temperature dependent permeability of SO$_2$, $^3$He atoms pass through the glass only when heated with an attached heating resistor, and can subsequently be ionized by an electron beam from a field-emission point. As indicated in Fig.~\ref{breitrabi_fig}, driving the hyperfine transitions requires microwaves of approximately $150~\si{\giga\hertz}$ and $4~\si{\giga\hertz}$. The former can enter the trap chamber through a window using an oversized waveguide, while the latter are irradiated using the shown spin-flip coils. 

In the Penning trap, the ion is confined radially by the homogeneous magnetic field along the $z$-axis and oscillates harmonically along the field lines with frequency $\nu_z$ due to the quadrupolar electrostatic potential created by the trap electrodes. The superposition of the magnetic and electrostatic fields leads to two eigenmotions in the radial plane: the modified cyclotron and the magnetron motion, with frequencies $\nu_+$ and $\nu_-$, respectively.
From the measured eigenfrequencies the free cyclotron frequency $\nu_c$ is calculated via the so-called invariance theorem $\nu_c=\sqrt{\nu_+^2+\nu_z^2+\nu_-^2}$,
where eigenfrequency shifts caused by trap misalignment and ellipticity cancel~\cite{invariance}.
In order to measure the motional eigenfrequencies, a superconducting tank circuit is attached to one trap electrode and converts the image current induced by the axial motion of the ion into a detectable voltage "dip" signal~\cite{dip}. The two radial motions do not couple directly to the resonator but are thermalized and detected using radio-frequency sideband coupling~\cite{doubledip}.

In the AT, the continuous Stern–Gerlach effect~\cite{sterngerlach} is utilized to detect changes of the hyperfine state. The quadratic inhomogeneity $B_2$ created by the ferromagnetic electrode leads to an additional term $\Delta\Phi(z)= -\mu B_2z^2$ to the potential along the $z$-axis, coupling the ion's magnetic moment $\mu$ to the axial frequency $\nu_z$. Thus, a spin-flip which changes the ion's magnetic moment by $\Delta\mu$ results in a shift of the axial frequency
\begin{equation}
    \Delta\nu_z= \frac{1}{2\pi^2\nu_z}\frac{B_2\Delta\mu}{m_{^3\textnormal{He}^+}}.
    \label{signal}
\end{equation}
As shown in the Breit-Rabi diagram (Fig.~\ref{breitrabi_fig}), the electronic transitions $\ket{1}\leftrightarrow\ket{3}$ and $\ket{2}\leftrightarrow\ket{4}$ or the nuclear transitions $\ket{1}\leftrightarrow\ket{2}$ and $\ket{3}\leftrightarrow\ket{4}$ effectively correspond to an electronic or nuclear spin-flip. 
An electronic spin-flip can be detected via a $\Delta\nu_z=\pm22~\si{\hertz}$ jump of the axial frequency as depicted in Fig.~\ref{fig_traps}B. 
A nuclear spin-flip, by contrast, causes a signal $\Delta\nu_z$ which is smaller by three orders of magnitude in the same magnetic inhomogeneity, since $\mu_e/\mu_I\approx1000$. 
Due to the inverse scaling of $\Delta\nu_z$ with the ion mass (see Eq.~\eqref{signal}), directly  detecting nuclear spin-flips over the background of axial frequency noise \cite{axnoise} is possible only for small masses and has so far been demonstrated only for protons and anti-protons~\cite{300ppt,anti2017}. Compared to a proton, $^3$He$^{2+}$ has a larger mass and smaller spin magnetic moment so that the signal indicating a spin-flip is smaller by a factor of four and not detectable unless the axial frequency noise is reduced significantly, for example through sympathetic laser-cooling~\cite{lasercooling}.
However, in case of $^3$He$^{+}$ a novel method can be employed which deduces the nuclear spin state from more easily detectable electronic transitions. 
If the ion is in hyperfine state $\ket{1}$ or $\ket{3}$ the nuclear spin state is $\ket{\uparrow}$, while states $\ket{2}$ and $\ket{4}$ imply that the nuclear spin state is $\ket{\downarrow}$, compare Fig.~\ref{breitrabi_fig}.
Thus, depending on the nuclear state only one of the two electronic transitions $\ket{1}\leftrightarrow\ket{3}$ and $\ket{2}\leftrightarrow\ket{4}$ can be driven. 
The nuclear state can therefore be found by exciting both electronic transitions alternatingly until a spin-flip occurs.

\begin{figure}
 \includegraphics[width=136mm]{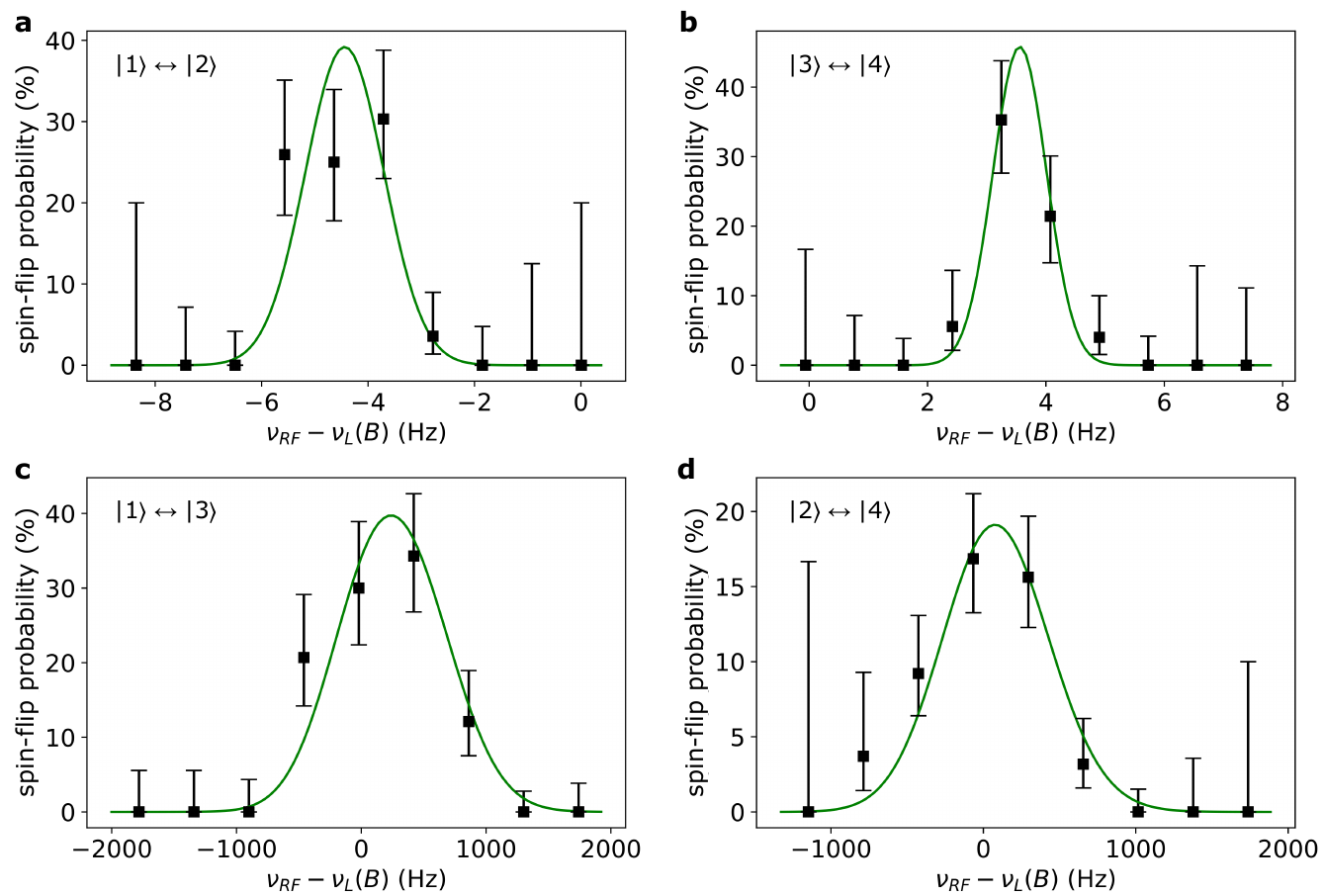}
    \caption{Exemplary resonance curves for each of the four hyperfine transitions. The $x$-axis is the difference of the frequency at which the spin-flip was driven and the expected resonance frequency at the simultaneously measured $B$-field, assuming the Breit-Rabi equation with the theoretically calculated parameters. The green line is calculated from a maximum likelihood analysis assuming a Gaussian lineshape. (a) and (b): nuclear spin-transitions $\ket{1}\leftrightarrow\ket{2}$ and $\ket{3}\leftrightarrow\ket{4}$, where the names of the states relate to the Breit-Rabi diagram in Fig.~\ref{breitrabi_fig}. (c) and (d): electron spin-transitions $\ket{1}\leftrightarrow\ket{3}$ and $\ket{2}\leftrightarrow\ket{4}$. All error bars correspond to the $1\sigma$ confidence interval (68\%).}
    \label{fig_resonances}
\end{figure}
Both the nuclear and electronic resonances were measured several times for different microwave powers and exemplary resonance curves are shown in Fig.~\ref{fig_resonances}. 
The parameters $g_e$, $g_I'$ and $E_{\text{HFS}}$ are extracted by a maximum likelihood analysis assuming a Gaussian lineshape. The systematic uncertainty imposed by non-analytical lineshape modifications of the resonance curves (Tab.~\ref{sys.shifts}) is calculated from the deviation of a Gaussian lineshape from the two asymmetric lineshapes derived in~\cite{geo_line, verdu}, which take the residual magnetic field inhomogeneity in the PT into account (see supplementary material). The final values include only measurements with small microwave powers where the results are lineshape model independent. They are corrected for the systematic shifts due to electrostatic and magnetic field imperfections, the axial dip fit, relativistic mass increase and the image charge induced in the trap electrodes~\cite{electronm,300ppt,anti2017,imagecharge,jk_shifts}, see Tab.~\ref{sys.shifts}.
The two parameters $g_I'$ and $E_{\text{HFS}}$ only have a weak dependence on the electron \textit{g}-factor and are determined by combining one resonance of each nuclear transition in one fit while leaving $g_e$ fixed to the theoretical value. Similarly, the electron \textit{g}-factor is fitted with a fixed value for the two nuclear parameters $g_I'$ and $E_{\text{HFS}}$ on which the electronic transition frequencies depend only weakly. In each case, changing the fixed parameter by $3\sigma$ leads to a shift of the result which is more than two orders of magnitude smaller than the statistical uncertainty.

\begin{figure*}
\includegraphics[width=89mm]{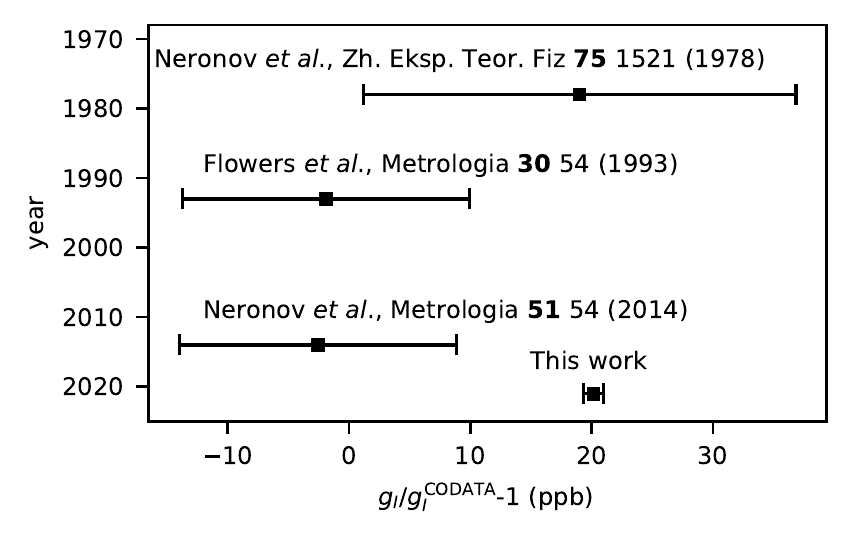}
\caption{History of $^3$He nuclear $g$-factor determinations. Comparison of previous measurements of the bare nuclear $g$-factor $g_I$ of $^3$He and the value given in this work. All previous results were derived from comparisons of the NMR frequency of $^3$He to that of water or molecular hydrogen. All error bars correspond to the $1\sigma$ confidence interval (68\%).}
\label{fig_history}
\end{figure*}

The result for the shielded nuclear \textit{g}-factor $g'_I=-4.255\, 099\, 606\, 9(30)_{stat}(17)_{sys}$ is used to calculate the $g$-factor of the bare nucleus $g_I=g_I'/(1-\sigma_{^3\textnormal{He}^+})=-4.255\, 250\, 699\, 7(30)_{stat}(17)_{sys}(1)_{theo}$. The latter uncertainty is due to the theoretical value for the diamagnetic shielding $\sigma_{^3\textnormal{He}^+}$. The shielded magnetic moment that provides the calibration of $^3$He NMR probes $\mu_{^3\textnormal{He}}=\mu_N/2\cdot g_I(1-\sigma_{^3\textnormal{He}})$ then follows by inserting the calculated shielding factor $1-\sigma_{^3\textnormal{He}}$ of atomic $^3$He \cite{sighe} and the nuclear magneton $\mu_N$ \cite{CODATA2018}. The latter two values have a relative uncertainty of $1\cdot10^{-10}$ and $3\cdot10^{-10}$ and the result $\mu_{^3\textnormal{He}}=-16.217\,050\,033(14)~\si{\mega\hertz\per\tesla}$ is one order of magnitude more precise than the most precise indirect determination \cite{neronov14}. This is the first stand-alone calibration for $^3$He probes, applicable for example in the muon $g-2$ experiments~\cite{muon,japan} which currently rely on water NMR probes. Our value for $g_I$ is compared to previous indirect determinations in Fig.~\ref{fig_history}. The relative deviation of $22$~p.p.b. from the most precise indirect result corresponds to three times the resonance linewidth or likewise a relative shift of the measured $B$-field by $10^{-8}$. Such a systematic shift in the magnetic field measurement can be excluded due to the agreement within 1$\sigma$ of the theoretical electron \textit{g}-factor $g_e^\text{theo}$, see above, and the experimental result $g_e^\text{exp}=-2.002\, 177\, 415\, 79(34)_{stat}(30)_{sys}$, which was measured more than one order of magnitude more precisely than $10^{-8}$. The indirect determinations of $g_I$ assume the shielding in water at $25~\si{\degree}$C of $\sigma_{\textnormal{H}_2\textnormal{O}}=25.691(11)\cdot10^{-6}$ \cite{CODATA2018} and the measured NMR frequency ratio $\nu'_{\textnormal{H}_2\textnormal{O}}/\nu'_{^3\textnormal{He}}$. Accordingly, combining this frequency ratio \cite{flowers93} with our result for $g_I$ yields a deviating shielding in water of $\sigma_{\textnormal{H}_2\textnormal{O}}=25.6689(45)\cdot10^{-6}$, using
\begin{equation}
     \frac{1-\sigma_{\textnormal{H}_2\textnormal{O}}}{1-\sigma_{^3\textnormal{He}}}=\frac{\nu'_{\textnormal{H}_2\textnormal{O}}}{\nu'_{^3\textnormal{He}}}\frac{\abs{g_I}}{g_{p}}.
    \label{shratio}
\end{equation}
Here, $g_p$ is the proton $g$-factor~\cite{300ppt}. This result corresponds to a relative uncertainty of 4.5 p.p.b. for the shielded magnetic moment in water $\mu_{\textnormal{H}_2\textnormal{O}}=\mu_N/2\cdot g_p(1-\sigma_{\textnormal{H}_2\textnormal{O}})$, limited by the uncertainty of the frequency ratio measurement.

The difference between our theoretically calculated $E_\textnormal{HFS}^{\rm theo}$, given above, and the much more accurate experimental value of $E_\textnormal{HFS}^{\rm exp}=-8\, 665\, 649\, 865.77(26)_{stat}(1)_{sys}~\si{\hertz}$ is 6~parts per million (p.p.m.). 
In a previous theoretical work, the discrepancy is 46 p.p.m. \cite{Friar2005}. In Ref.~\cite{Karshenboim2002}, a difference of 222 p.p.m. between the QED prediction and the experimental value is taken as an estimate of contributions to HFS due to nuclear effects.
The experimental result $E_\textnormal{HFS}^{\rm exp}$ is in agreement with the previous most precise measurement $-8\, 665\, 649\, 867(10)~\si{\hertz}$~\cite{EHFS_1969}, while improving the precision by two orders of magnitude. It is used to extract the Zemach radius $r_Z=2.608(24)~\si{\femto\meter}$, as described in the supplementary material, which differs by $2.8~\sigma$ from $r_Z=2.528(16)$, previously determined from electron scattering data \cite{Sick2014}.

In the future, improved measurements are possible by first reducing the magnetic inhomogeneity of the precision trap, which reduces the resonance line widths as well as systematic effects on the resonance lineshape, and second introducing phase-sensitive detection methods for more precise magnetic field measurements \mbox{\cite{deuteronmass}}. In addition, the measurement method described here can be applied to determine the nuclear magnetic moment of other hydrogen-like ions which are too heavy for direct nuclear spin-flip detection via the Stern-Gerlach effect. We note that He$^+$ is the only one-electron ion where uncertainties arising from nuclear structure are small enough to additionally enable a competitive determination of $\alpha$ \cite{Zatorski2017}, provided that the experimental uncertainty of $g_{e}$ can be decreased in future by orders of magnitude. As a next step, the magnetic moment of the bare $^3$He$^{2+}$ nucleus can be measured directly in a Penning trap with a relative precision on the order of 1 p.p.b. or better by implementing sympathetic laser cooling \cite{3he2p_trap}.


\begin{table*}[]
\caption{Corrections to the nuclear \textit{g}-factor, electron \textit{g}-factor and zero-field hyperfine splitting due to systematic effects.}
\begin{center}
  \begin{tabular}{c|l|l|l}
    effect & $\Delta g_I'/g_I'$~$(10^{-10})$ & $\Delta g_e/g_e$~$(10^{-10})$ & $\Delta E_{\text{HFS}}/E_{\text{HFS}}$~$(10^{-12})$\\ \hline
    relativistic & -0.33(2)&-0.21(1) &-0.084(4)\\
    image charge & -0.514(3)&-0.321(2) &-0.128(1)\\
    electrostatic anharmonicity &-0.03(5) &-0.02(3) &-0.01(1)\\
    magnetic inhomogeneity &\phantom{-}0.17(2)&\phantom{-}0.11(1) &\phantom{-}0.044(4)\\
    axial dip-fit &\phantom{-}0(0.5) &\phantom{-}0(0.3) &\phantom{-}0(0.1)\\
    resonance lineshape &\phantom{-}0(4) &\phantom{-}0(1.5) &\phantom{-}0(1)\\\hline
    $\Sigma$ & -0.7(4.0) &-0.4(1.5) &-0.2(1.1)
  \end{tabular}
\end{center}

\label{sys.shifts}
\end{table*}
\section*{Acknowledgments}
This work is part of and funded by the Max Planck Society and RIKEN. Furthermore this project has received funding
from the European
Research Council (ERC) under the European Union’s
Horizon 2020 research and innovation programme under
grant agreement No. 832848 - FunI and we acknowledge
funding and support by the International Max Planck Research School for Precision Tests of Fundamental Symmetries (IMPRS-PTFS) and by the Max Planck RIKEN PTB Center for Time, Constants and Fundamental
Symmetries. 
We acknowledge helpful discussions with T. Chupp, T. Mibe, K. Shimomura, K. Sasaki, W. Heil, P. Blümler, H. Busemann and M. Moutet.
\section*{Author Contributions}
A.M., A.S., S.D. and M.M. performed the measurements and B.S., Z.H., N.S.O. and I.A.V. carried out the QED calculations. The manuscript was written by A.S., A.M., S.U., K.B., Z.H., B.S., N.S.O. and I.A.V. and discussed among and approved by all co-authors.
\section*{Competing interests}
The authors declare no competing interests.
\section*{Data Availability}
The datasets generated and analyzed during this study are available from the corresponding author on request.
\section*{Code Availability}
The code used during this study is available from the corresponding author on request.

\end{document}